\documentclass[STIX1COL]{WileyNJD-v2}

\usepackage{graphicx}

\makeatletter

\providecommand{\tabularnewline}{\\}

\usepackage{url}

\makeatother

\usepackage{listings}
\begin{document}

\articletype{Research Article} 

\received{15 December, 2020} 
\revised{3 March, 2021} 
\revised{24 May, 2021} 

\author[1]{Yaniv Rubinpur}
\author[1]{Sivan Toledo*}

\address[1]{\orgdiv{Blavatnik School of Computer Science}, \orgname{Tel Aviv University}, \orgaddress{\country{Israel}}}

\corres{*Sivan Toledo, \email{stoledo@tau.ac.il}}
\presentaddress{Blavatnik School of Computer Science, Tel Aviv 69978, Israel} 


\title{Signal Processing for a Reverse-GPS Wildlife Tracking System: CPU
and GPU Implementation Experiences}

\abstract[Summary]{We present robust high-performance implementations of signal-processing
tasks performed by a high-throughput wildlife tracking system called
ATLAS. The system tracks radio transmitters attached to wild animals
by estimating the time of arrival of radio packets to multiple receivers
(base stations). Time-of-arrival estimation of wideband radio signals
is computationally expensive, especially in acquisition mode (when
the time of transmission is not known, not even approximately). These
computations are a bottleneck that limits the throughput of the system.
We developed a sequential high-performance CPU implementation of the
computations a few years back, and more recencely a GPU implementation.
Both strive to balance performance with simplicity, maintainability,
and development effort, as most real-world codes do. The paper reports
on the two implementations and carefully evaluates their performance.
The evaluations indicates that the GPU implementation dramatically
improves performance and power-performance relative to the sequential
CPU implementation running on a desktop CPU typical of the computers
in current base stations. Performance improves by more than 50X on
a high-end GPU and more than 4X with a GPU platform that consumes
almost 5 times \emph{less} power than the CPU platform. Performance-per-Watt
ratios also improve (by more than 16X), and so do the price-performance
ratios.}

\maketitle

\section{Introduction}

ATLAS is a reverse-GPS wildlife tracking system, targeting mostly
regional movement patterns (within an area spanning kilometers to
tens of kilometers) and small animals, including small birds and bats~\cite{atlas-sw-eng,atlas-accuracy}.
ATLAS is a mature collaborative research effort: 6 systems have been
set up and are operating in 5 countries on 3 continents. The first
system has been operating for about 6 years almost continuously and
has produced ground-breaking research in Ecology~\cite{OwlMicrobial,ATLASScience}. 

ATLAS tracks wild animals using miniature radio-frequency (RF) transmitting
tags attached to the animals~\cite{TagsEDERC2014,vildehaye-wp}.
The transmissions are received by ATLAS base stations that include
a sampling radio receiver and a computer running Linux or Windows.
The computer processes RF samples to detect transmissions from tags
and to estimate the time of arrival (ToA) of the transmissions. It
reports the reception times to a server via an internet connection,
usually cellular. The server estimates the location of a tag from
ToA reports of the same transmission by different base stations~\cite{atlas-accuracy}. 

The signal processing that ATLAS base stations performs is computationally
demanding and is one of the main limiting factors of the throughput
of the system (the number of tags that it can track and the number
of localizations per second that it can produce).

We developed a high-performance sequential (single-threaded) implementation
of the algorithms a few years ago. More recently, we developed a new
implementation\footnote{The current CPU and GPU versions of the code are available, along
with the data files requires to run the code, at \url{http://www.tau.ac.il/~stoledo/Tools/atlas-dsp-heteropar2020.zip}.} designed to effectively exploit graphical processing units (GPUs).
Our aim in developing this implementation was to significantly improve
the throughput of the system and to reduce the power consumption of
base stations. Reduced power consumption reduces the cost and complexity
of base stations that rely on solar and wind energy harvesting, such
as those deployed in the shallow Wadden sea; it is not particularly
important in base stations connected to the power grid. High throughput
is useful in most base stations.

Our objectives for both implementations included not only high performance
and energy efficiency, but also maintainability and the scope of the
development effort. Our implementations balance these concerns, as
any real-world production code should. They represent the outcome
of real-world research and development, not of a purely academic exercise.

We evaluate the performance of both the (slightly improved) CPU code
and the new GPU code on real recorded data. The evaluations, performed
on two CPU platforms and on three GPU platforms, show dramatic improvements
relative to our baseline, a high-end desktop CPU that is typical of
the computers in current base stations. The improvements are both
in terms of absolute performance (more than 50X with a high-end GPU
and more than 4X with a GPU platform that consumes almost 5 times
\emph{less} power than the CPU platform), in terms of performance-per-Watt
ratios (more than 16X), and in terms of price-performance ratios.
However, because we did not attempt to achieve top multicore performance
on CPUs, these results should not be taken as fair comparisons of
the hardware platforms; they are meant mainly to demonstrate the level
of performance that is achievable on such tasks on GPUs using a single-threaded
scheduler coupled with GPU data parallel tasks.

The rest of this paper is organized as follows. Section~\ref{sec:Background}
provides background material required to understand our contributions.
It describes how ATLAS base stations operate, how their scheduler
operates, and the basics of GPU programming with CUDA. Section~\ref{sec:DSP-algs}
describes the main signal processing algorithms that ATLAS base stations
use to detect tag transmissions and to estimate their arrival times.
The CPU and GPU implementations of these algorithms are described
in Section~\ref{sec:High-Performance-Implementation}. The results
of our evaluation of the performance and power consumption of the
two implementation on a variety of platforms are described in Section~\ref{sec:Experimental-Evaluation}.
The new GPU implementation is already deployed in the field; experiences
from this deployment are described in Section~\ref{sec:Deployment}.
Section~\ref{sec:Related-Work} describes related work, and Section~\ref{sec:Discussion-and-Conclusions}
discusses our results and presents our conclusions from this research.

A preliminary version of this paper has been published in the proceedings
of HeteroPar 2020~\cite{rubinpur2020highperformance}.

\section{Background}

\label{sec:Background}ATLAS tags transmit a fixed unique pseudorandom
packet every second, 2~s, 4~s, or 8~s. The packets are 8192-bit
long and the bitrate is around 1~Mb/s. The data is frequency modulated
(FSK); ATLAS can also use phase modulation (PSK; see~\cite{modulation}
for details), but in this paper we focus on signal processing for
frequency modulation, which is what almost all the deployed tags use.
The sampling receiver in each base station sends a continuous stream
of complex RF samples, usually at 8 or 8.33~Ms/s, to a computer.
The samples are placed in a circular buffer. A high-level scheduler
repeatedly extracts a block of samples from the buffer and processes
it. The size of the circular buffer allows for processing delays of
more than 10~s; this simplifies the scheduler and the signal-processing
code considerably relative to in-order stream processing with hard
deadlines.

The signal processing aims to detect whether packets from specific
tags appear in the block, to estimate the precise (sub-sample) time
of arrival (ToA) of each packet, to estimate the (relative) power
of the packet, and to estimate a signal-to-noise ratio that is correlated
with the variance of the ToA estimate. This data is sent to a server
that estimates the locations of the tags~\cite{atlas-accuracy}.

\subsection{The High-Level Scheduler}

ATLAS base stations use a high-level scheduler, implemented in Java.
The scheduler creates two kinds of tasks for the signal-processing
code. \emph{Searching-mode} (acquisition-mode) tasks process blocks
of 100~ms and try to detect packets from all the tags that have not
been detected in the past few minutes. This set of tags is called
the \emph{searching queue}. It can consist of over 100 tags. Since
all tags transmit on one or two frequencies, the FSK demodulation
step is performed only once or twice per block of samples, but the
number of pseudorandom codes that must be correlated with the demodulated
signals can be large. \emph{Tracking-mode }tasks aim to detect an
8~ms packet from one particular tag in a block of about 12~ms of
samples. These tasks perform demodulation and correlate the demodulated
signal with one pseudorandom code. 

Normally, the scheduler allocates 50\% of the processor's time to
searching and 50\% to tracking, in an amortized sense, simply to avoid
starvation of one of the tasks. If one of the queues is empty, all
the processing resources are devoted to the other queue. 

The scheduler is sequential; it generates one task at a time and performs
it to completion, devoting to it all the cores except for one that
handles incoming samples. this simplifies its algorithms but places
all the responsibility to efficiently utilize multiple cores to the
signal-processing code.

\begin{figure}
\begin{centering}
\includegraphics{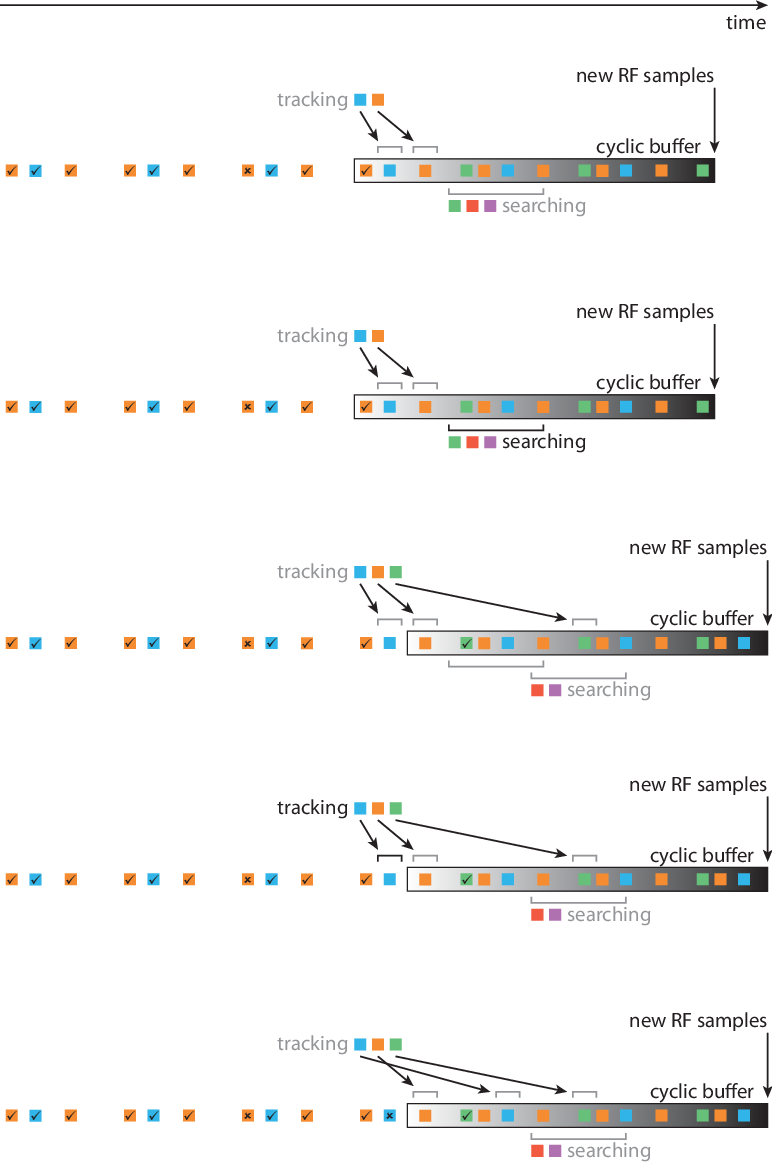}
\par\end{centering}
\caption{\label{fig:timeline}An illustration of the actions of the scheduler.
Rows represent actions (steps) that the scheduler takes over time.
Within each row, events are ordered from left to right. Each colored
square represents a received transmission from a tag; the color represents
the tag. The horizontal span of the square represents a span of time
(samples) in which the transmission was received. See text for further
explanation of the figure.}

\end{figure}
The behavior of the scheduler is illustrated in Figure~\ref{fig:timeline}.
Each row in the figure depicts the state of the scheduler at a particular
time and in most rows, also a decision made by the scheduler. Within
each row, events are ordered from left to right. Each colored square
represents a received transmission from a tag; the color represents
the tag. The horizontal span of the square represents a span of time
(samples) in which the transmission was received. In the top line,
the base station has already successfully detected the six transmissions
of the orange tag and the three of the blue tag shown to the left
of the cyclic buffer. The check marks indicate that these transmissions
were successfully detected. The tracking queue consists of the orange
and blue tags, and the scheduler correctly predicts their first unprocessed
transmissions, which are currently stored in the buffer. The buffer
also contains one transmission of the orange tag that is still stored
in the buffer, as well as additional transmissions of the orange,
blue, and purple tags. The RF samples are stored at the front (right
end) of the buffer and when that happens, old samples from the read
(left end of the buffer) are discarded.

The second row represents the next action of the scheduler: it decides
to serve the searching queue. It processes the range of samples shown
by the bracket below the buffer. This range contains transmissions
of the orange and blue tags, which the searching tag does not search
for, and a transmission of the purple tag, which the task does search
for. The third row shows the outcome of searching. The transmission
of the green tag was detected, so it is checked. This caused the green
tag to move from the searching to the tracking queue. The scheduler
correctly predicts its next transmission time. Also, searching moves
forward, with a 10~ms overlap with the previous span, to ensure that
transmissions are fully contained in the span of some searching task
(indeed, a transmission of the orange tag was partially contained
in the first searching task). While the searching task was processed,
new samples have arrived in the buffer and old ones discarded.

The fourth row indicates the next action of the scheduler. It decides
to serve the tracking queue, and within that queue it processes the
oldest transmission. In the figure, it is a transmission of the blue
tag, which is no longer in the buffer. The bottom row shows the outcome
of the scheduler's action: the missed transmission is marked as such
(with an 'x') and the prediction of the next transmission of the blue
tag is advanced by one inter-transmission period, to a transmission
that is still in the buffer.

\subsection{Useful Performance Goals}

Performance improvements in the signal-processing code are useful
only up to a certain point, due to the structure of the ATLAS system.
This section documents these limits.

Tags that use FSK modulation create significant inter-tag interference
when the transmissions of two or more tags that transmit on the same
frequency overlap in time~\cite{modulation}. The interference depends
on the extent of the overlap. ATLAS systems have tracked over 60 such
tags simultaneously and successfully; at these numbers, there are
already some overlaps (the transmission times are not coordinated)
and much larger numbers, say 100 tags, are likely to degrade localization
accuracy significantly. Therefore, in a single-channel (single frequency)
system the number of tags that need to be simultaneously tracked is
around 60--100. However, ATLAS receivers sample 8~MHz or more, so
they are capable of processing at least 4 separate channels. It is
easy to configure the receivers to even higher bandwidth, at least
32~MHz. Multichannel ATLAS systems can track hundreds of tags simultaneously
if the signal processing is fast enough.

Searching performance affects how quickly a tag is detected once its
transmissions are physically received by a base station. Tags that
are too far from a base station cannot be detected. So are tags in
caves (common for bats) or underground burrows. Tags that are very
close to the ground are also often not detectable, due to interference
from ground reflections. When a tag whose transmission was undetectable
moves to a location from which the received signal is strong enough,
RF samples contain a representation of the transmission, but unless
these RF samples are searched, the tag is not detectable. If a base
station can search all the RF samples for, say, 25 tags, but the system
is trying to track 50 tags that cannot currently be detected, then
in expectation it will take two transmissions of a tag until it is
detected, because the base station only processes 50\% of the RF samples
for the tag. In ATLAS systems in which most of the tags are received
most of the time, searching performance is not important. But in systems
in which a large number of the active tags are usually not detectable,
searching performance reduces the lag until a tag is detected. The
ability to search for 50 or 100 tags in all the RF samples is certainly
useful for existing single or dual-channel ATLAS systems that track
bats that spend day time in caves.

\subsection{General-Purpose GPUs and CUDA}

Graphics cards (GPUs) that can run general-purpose code, sometimes
called GPGPUs, have emerged as effective accelerators of computationally-intensive
tasks~\cite{GPUsReshapeComputing}. This paper focuses on GPUs produced
by the market leader, NVIDIA. NVIDIA GPUs contains a large number
of simple cores (execution units) under the control of a smaller number
of instruction schedulers. In the Jetson TX2 GPU, for example, 256
cores are organized into \emph{warps} of 32 cores that are controlled
by a single instruction scheduler. The warps are organized into \emph{streaming
multiprocessors} (SMs; two in the TX2). All the cores in a warp perform
the same operation at the same time, so the code must exhibit a high
degree of data parallelism. Larger NVIDIA GPUs use the same basic
structure, but with different numbers of cores and SMs. Many NVIDIA
GPUs can only operate directly on data stored in the GPUs memory,
not in the computer's main memory. NVIDIA GPUs have a memory hierarchy
that includes a small block of so-called \emph{shared} memory that
is private to an SM; on the TX2, its size is 64~KB. Data-movement
engines called \emph{copy engines }in the GPU move data between main
memory and GPU memories and within GPU memories.

NVIDIA GPUs run programs written in CUDA, an extension of the C language.
CUDA programs consist of host code that runs on the CPU and invokes
\emph{GPU kernels }to perform data-parallel operations.  As explained
above, our code always separates the memory allocation from invocation
of kernels, to reduce overhead. We also reduce data movement between
CPU and GPU memory as much as possible, storing vectors that are reused,
like transformed codes, in GPU memory.

The CUDA implementation of a kernel that adds two vectors is extremely
simple.  
\begin{lstlisting}
__global__ void add(int *a, int *b, int *c) { 
  int index = blockDim.x * blockIdx.x + threadIdx.x;
  c[index] = a[index] + b[index]; 
}
\end{lstlisting}

Reductions (summations, maximum value in an array, etc) are quite
difficult to implement in CUDA. There are libraries that implement
reductions, but the use of these libraries does not allow a data parallel
operation to be fused with a reduction, which increases memory traffic.
A C++ source library called CUB~\cite{CUB} simplifies the implementation
of reductions and allow them to be fused with data parallel operations.
CUB uses shared memory to achieve high performance; it requires the
caller to allocate this memory, which our code does. Here is a simple
reduction implemented using CUB; it is typical of reductions in our
code.
\begin{lstlisting}
__global__ void maxInt(int* input, int* out)
{
    typedef BlockReduce<int, THREADS> BlockReduceT;     // a CUB reducer

    __shared__ typename BlockReduceT::TempStorage temp; // shared memory for CUB

    int index = blockDim.x * blockIdx.x + threadIdx.x

    // CUB computes the reduction over threads in each block
    float block_max = BlockReduceT(temp_storage).Reduce(input[index], cub::Max());

    // reduction over the blocks using a CUDA atomic primitive
    if(threadIdx.x == 0) {
        atomicMax(out, block_max);
    }
}
\end{lstlisting}

\section{Signal Processing in ATLAS}

\label{sec:DSP-algs}The signal processing that RF samples associated
with searching or tracking tasks is virtually identical. The samples
undergo the mathematical transformations described below. However,
the transformations are not applied naively, but in an optimized way
described in Section~\ref{sec:High-Performance-Implementation}.
We focus for clarity only on FSK; signal processing for phase-shift
keying (PSK) is described by Leshchenko and Toledo~\cite{modulation}.
The mathematical transformations are:
\begin{enumerate}
\item Conversion of the complex RF samples, residing in the cyclic buffer
and represented by pairs of 16-bit integers, to a single-precision
(\texttt{float}) complex vector $x$. 
\item The complex samples are usually multiplied element-wise by a complex
input vector $l$ representing a local-oscillator signal, to shift
the center frequency so that the spectrum of transmissions is centered
at zero. That is, we replace $x\leftarrow x\odot l$ (for all $i$,
$x_{i}\leftarrow x_{i}\cdot l_{i}$). 
\item Next, a bandpass FIR (finite impulse response) filter, represented
here by a circulant matrix that $H_{\text{BP}}$, is applied, to produce
$y\leftarrow H_{\text{BP}}x$. We use filters with 200 coefficients.
\item Two short (8 samples) matched filters are applied to $y$, one that
represent a single-bit (chip) period at the frequency representing
a $1$ symbol and one that represent a single-bit period at the frequency
that represents a $0$ symbol. We denote their outputs by $f_{1}=H_{1}y$
and $f_{0}=H_{0}y$. 
\item The vectors $f_{1}$ and $f_{0}$ are used to demodulate the transmission
in two different ways, with and without normalization,
\begin{eqnarray*}
d & = & \left(\left|f_{1}\right|-\left|f_{0}\right|\right)\oslash\left(\left|f_{0}\right|+\left|f_{0}\right|\right)\\
u & = & \left|f_{1}\right|-\left|f_{0}\right|
\end{eqnarray*}
(elementwise absolute value, elementwise subtraction and addition,
and elementwise division). These signals are real. 
\item The algorithm applies exactly the same steps to a \emph{replica} of
the transmission we are trying to detect, a synthetic noise-free zero-padded
signal $r^{(c)}$ that represents an FSK packet with the same modulation
parameters and a pseudo-random bit sequence $c$. The resulting demodulated
vector is denoted $d^{(c)}$; it is computed once and stored. The
discrete signal $r^{(c)}$ is padded so that the length of $d^{(c)}$
is identical to the lengths of $d$ and $u$. 
\item We cross-correlate $d$ with $d^{(c)}$. The cross correlation vector
is also real. 
\item We compute the value and location $j$ of the maximum of the absolute
value of the cross correlation vector, 
\[
j=\arg\max_{i}\left|\text{xcorr}(d,d^{(c)})\right|\;.
\]
The elements of $\text{xcorr}(d,d^{(c)})$ around $j$ are subsequently
interpolated to estimate the arrival time of the incoming signal.
We also compute quantities that are used to estimate the signal-to-noise
ratio (SNR) and the power of the signal. More specifically, assuming
that the nonzero part of $d^{(c)}$ spans its first $n$ elements,
we compute:
\begin{eqnarray*}
w_{c} & = & \sum_{i=0}^{n}d_{i}^{(c)}d_{i+j}\;,\\
q & = & \sum_{i=0}^{n}d_{i+j}^{2}\;,\text{ and }\\
p_{c} & = & \sum_{i=0}^{n}d_{i}^{(c)}u_{i+j}\;.
\end{eqnarray*}
For details on how power and SNR are estimated and how they are used,
see~\cite{modulation,rubinpur2020highperformance}. 
\end{enumerate}

\section{High-Performance Design and Implementations}

\label{sec:High-Performance-Implementation}Our implementations of
the computation described above are optimized. We use fast Fourier
transforms (FFTs) to reduce the operation counts. We also use high-performance
implementation principles in both the CPU implementation in C and
in the GPU implementation in CUDA. In most cases the principles are
applicable to both implementations; we highlight the differences when
this is not the case.

\subsection{Using Fast Fourier Transforms}

\begin{figure}
\begin{centering}
\includegraphics{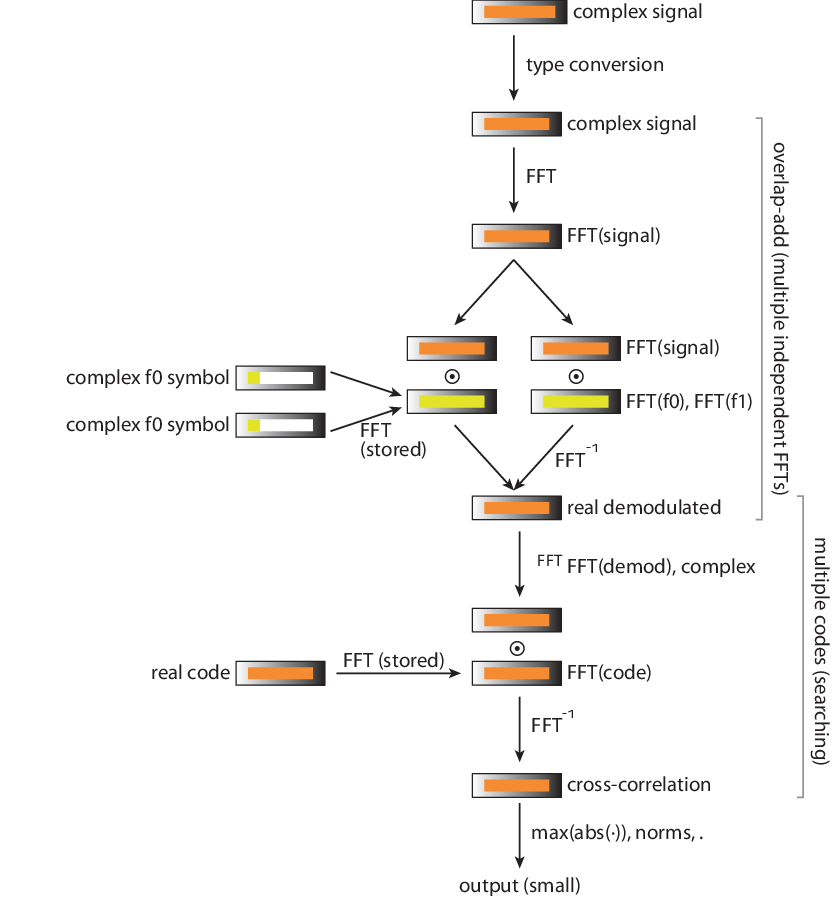}
\par\end{centering}
\caption{\label{fig:dsp}The signal processing data flow in ATLAS.}
\end{figure}
We use techniques that minimize the operation counts that the signal-processing
building-blocks perform. 

In particular, we use the fast-Fourier transform (FFT) to compute
cross-correlation and to apply FIR filters with many coefficients.
To compute cross correlation, we use the identity $\text{xcorr}(d,d^{(c)})=\text{ifft}(\text{fft}(d)\odot\text{fft}(d^{(c)}))$.
We compose FIR filters that are applied in a sequence ($H_{\text{BP}}H_{1}$
and $H_{\text{BP}}H_{0}$) and we use FFTs to apply long FIR filters
(filters with many coefficients). The formula is similar, except that
the filters are naturally expressed using a convolution, not a cross
correlation. 

We use the overlap-add method to apply medium-length filters and cross
correlations. This reduces the operation count from $\Theta(m\log m)$
to $\Theta(m\log n)$ when applying a filter of length $n$ to a block
of $m$ RF samples. 

We pad input lengths to lengths that are a product of small integers,
usually 2, 3, and 5; this ensures that applying FFT is as inexpensive
as possible. 

The actual data flow in the code is shown in Figure~\ref{fig:dsp}.
Type conversion and demodulation is done once on the RF samples associated
with each task, both searching and tracking. The demodulation takes
advantage of the overlap-add method. The cross-correlation with the
demodulated signal is done once in tracking tasks (with the code that
we are tracking) and multiple times on the same demodulated signal
in searching tasks, once per code.

\subsection{Memorization and Planning}

The implementations allocate arrays when they are needed and reuse
them aggressively. In general, they are never released. For example,
demodulation of a block of RF samples of a certain size is always
done using the same temporary arrays; for other sizes, we use other
arrays. This reduces memory-allocation overheads. Allocated arrays
are aligned on cache-line boundaries in the CPU implementation and
are allocated in GPU memory in the GPU implementation.

Auxiliary vector, like $\text{fft}(d^{(c)})$, are computed when needed
and stored indefinitely, to avoid recomputation. 

Allocating auxiliary arrays of a given size and reusing them enables
\emph{preplanning} of all the FFT calls. We use comprehensive high-performance
FFT libraries to compute FFTs. On CPUs, we use FFTW~\cite{FFTW05};
on GPUs, we use NVIDIA's cuFFT. Both FFTW and cuFFT requires calls
to be \emph{planned} in order to achieve high performance. An FFT
plan refers to specific arrays at fixed (virtual) addresses, to allow
optimization based on cache lines and similar hardware boundaries.
Therefore, for every pair of arrays that are used as input and output
of FFT calls, we plan the corresponding FFT operation and retain in
memory both the plan and the input and output arrays.

\subsection{Reducing Communication and Batching}

Loops are aggressively fused in the CPU implementation and kernels
are aggressively fused in the GPU implementation. This reduces data
movement (e.g., cache misses) and allows elimination of some temporary
arrays.

In the CPU implementation, we also batch cross correlation operations:
a single call to FFTW computes many cross-correlation vectors. This
exposes ``embarrassing'' parallelism (completely independent operations)
that FFTW should be able to easily exploit, at least in principle.
Batching is possible during extend-add operations on single vectors,
and also when cross correlating a single demodulated signal with many
codes.

\subsection{CUDA Implementation}

Here is how our CUDA host code implements the data type conversion
and demodulation (the computation of $d$ and $u$ from the samples
$x$). The code is slightly simplified, but not by much. The vectors
\texttt{fd\_f0} and \texttt{fd\_f1} contain the transformed $H_{\text{BP}}H_{0}$
and $H_{\text{BP}}H_{1}$; they are preallocated, precomputed, and
stored on GPU memory.
\begin{lstlisting}
convertAndPad<<<BLOCKS,THREADS>>>(d_rf_samples, params, d_power_out, d_padded);

cufftExecC2C(plan_forward, d_padded, d_padded, CUFFT_FORWARD);

elementwiseMult<<<BLOCKS,THREADS>>>(d_padded, fd_f0, fd_f1, d_f0, d_f1);

cufftExecC2C(plan_backward, d_f0, d_f0, CUFFT_INVERSE);
cufftExecC2C(plan_backward, d_f1, d_f1, CUFFT_INVERSE);

elementwiseDemod<<<BLOCKS,THREADS>>>(d_f0, d_f1, demod_normalized, demod_unnormalized);
\end{lstlisting}

We use CUB to implement reductions, because it allows us to perform
multiple reductions in one pass over the data and to fuse reductions
with data parallel operations. We preallocate shared (fast) GPU memory
for CUB. 

In kernels that do not use CUB we do not use shared memory because
they implement low data-reuse data-parallel operations over large
vectors. The cuFFT library might also use shared memory, but if it
does, it allocates it internally.

\section{Experimental Evaluation}

\label{sec:Experimental-Evaluation}This section presents our experimental
evaluation of the effectiveness of GPUs for our task, in terms of
both performance and energy efficiency.

\subsection{Methodology (Test Data)}

To test the codes, we modified the CPU-based DSP C code so that it
stores all its inputs and outputs in files. We then ran the ATLAS
base station code in an ad-hoc mode (that is, not as part of a localization
system) on a computer connected to a USRP B210 sampling radio and
configured the base station to detect a tag that was present in the
room. This produced files that contained the RF samples that were
processed in both searching and tracking mode, inputs that represent
filter coefficients and the signal to correlate with, and the outputs
of the signal-processing algorithms.

Next, we wrote a C program that reads these files, calls the signal
processing routines on the recorded data, measures their running time
and optionally the power consumption of the computer and its components,
and stores the results in files. The program can use the recordings
in both single-code single-RF-window mode and in batch mode that processes
many codes in one call. The former is typical of tracking mode and
the latter of searching mode. The program checks that the returned
results are identical, up to numerical rounding errors, to those returned
by the full base station run that detected the tag correctly. This
ensures that all the results that we report represent correct executions
of the algorithms. The code then stores the running times and the
power measurements, if made, to log files.

We also tested that the new CUDA-based code works correctly when called
from Java through the JNI interface and detects transmissions from
tags and their arrival times. This test was performed on the Jetson
TX2 computer described below and the same URSP B210 radio.

\subsection{Platforms}

We evaluated the code on several platforms using both the CPU code
and the GPU code.

Our baseline is a small form-factor desktop computer, representative
of those currently used in ATLAS base stations, with an Intel i7-8700T
CPU. This CPU has 6 physical cores running at clock frequencies between
2.4 and 4~GHz and thermal design power (TDP) of 35W. This CPU was
launched in Q2 2018 and is fabricated in a 14~nm process. The computer
ran Linux kernel version~5.3. We compiled the code using GCC version~7.5.
Both our code and FFTW version 3.3.8 were compiled using the optimization
options that are built into FFTW. The code that was produced ran slightly
faster than code compiled with only \texttt{-O3 -mtune=native}. 

Our main target is a low-power Jetson TX2 computer~\cite{TX2HiddenDetails,TX2Blog},
which has a 256-core NVIDIA Pascal GPU, four ARM Cortex-A57 cores
and two ARM Denver2 cores, launched in Q2 2017 using a 16~nm process.
The Cortex-A57 cores were designed by ARM and the Denver2 cores were
designed by NVIDIA for higher single-threaded performance; both use
the same 64-bit ARMv8 instruction set. It also has 8~GB of memory
that both the CPU and GPU can access, with 59.7~GB/s memory bandwidth.
The TX2 ran Linux kernel 4.9.140-tegra. We used nvcc version~10.0.326,
CUDA library 10.0.130, gcc~7.4.0, and FFTW 3.5.7. CUB version~1.8.0
was used on all platforms.

We measured power consumption on the TX2 using two ina3221 current
sensors built into the TX2 module and a third built into the motherboard~\cite{JetsonLinux}.
Each sensor senses current on three different rails, and all the measurements
are available by reading special files exposed by the driver under
\url{/sys/bus/i2c/drivers/ina3221x}. The values that we report are
the maximum value observed during the computation.

The power-vs-performance profile of the TX2 can be adjusted by turning
cores on or off and by changing their clock frequency. NVIDIA defined
several standard modes, which we use below in our tests. Table~\ref{tab:tx2-modes}
describes these modes. The nominal TDP of the TX2 ranges from 7.5~W
for the highest power efficiency mode, to 15~W for the highest performance
modes. Both the TX2 module and the motherboards include power sensors
that we use to measure the power consumption directly in our tests.

\begin{table}[tb]
\centering{}\caption{\label{tab:tx2-modes}Standard power modes on the Jetson TX2.}
\begin{tabular}{lccc}
Mode Name & Denver2 Cores & A57 Cores & GPU Frequency\tabularnewline
\hline 
Max-Q & --- & $4\times1.2$~GHz & $0.85$~GHz\tabularnewline
Max-P All & $2\times1.4$~GHz & $4\times1.4$~GHz & $1.12$~GHz\tabularnewline
Max-P ARM & --- & $4\times2.0$~GHz & $1.12$~GHz\tabularnewline
Max-P Denver & $2\times2.0$~GHz & --- & $1.12$~GHz\tabularnewline
Max-N & $2\times2.0$~GHz & $4\times2.0$~GHz & $1.30$~GHz\tabularnewline
\end{tabular}
\end{table}

We also ran the GPU code on two additional platforms. One is an NVIDIA
GeForce 1050~GTX GPU. This GPU uses the Pascal architecture, 640
CUDA cores running at 1.455~GHz, and 2~GB of RAM. The TDP is 75~W.
It was plugged into a desktop running Windows~10 with a quad-core
Intel i5-6500 CPU; we used CUDA 10.1, nvcc version 10.1.168, Microsoft's
C++ compiler (cl) version 19.00.24210 for x64. The last GPU platform
that we used is an NVIDIA Titan Xp GPU. This GPU also uses the Pascal
architecture and has 3840 CUDA cores running at 1.582~GHz. It has
12~GB of memory and a high-bandwidth memory interface. The thermal
design power is 250~W. It was plugged into a server with a 10-core
Intel Xeon Silver 4114 CPU running Linux. We used CUDA and nvcc 10.0.130
and gcc 4.9.2. 

\subsection{Results}

\begin{figure}[tb]
\includegraphics[width=0.47\textwidth]{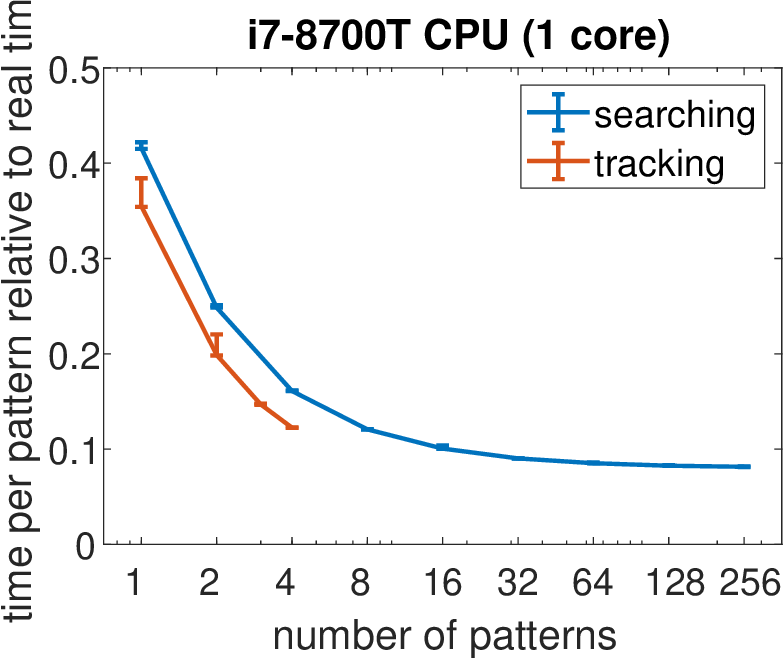}\hfill{}\includegraphics[width=0.47\textwidth]{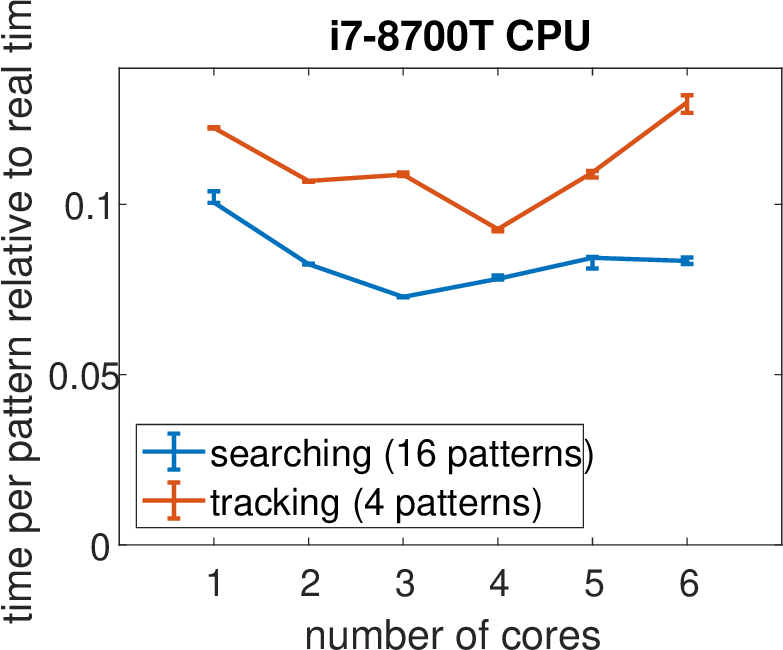}

\caption{\label{fig:i7-8700T}The performance of the DSP code on one CPU core
(left) and its speedup on multiple cores. The vertical bars show the
miniumum and maximum values over 10 experiments, and the actual data
points are median results of the 10 experiments. The number of RF
windows is 10 in the searching experiments and 100 in the searching
experiments. }
\end{figure}
Figure~\ref{fig:i7-8700T} shows the performance of our C implementation
on the baseline platform, which has an Intel i7-8700T CPU. We present
the performance in terms of the ratio of processing time per pattern
relative to the length of the RF window. That is, if the code takes
1~s to process one 100-ms window of RF samples and to correlate the
demodulated signal with 16 different code patterns, then we report
the performance as $(1/16)/0.1=0.625$. A ratio of $1$ implies that
the base station can search for one tag continuously, that searching
for 2 tags would drop 50\% of the RF samples, and so on. A ratio of
$0.1$ implies that the station can search continuously for 10 tags
without dropping any RF sample, and so on. Lower is better.

The results on one core (Figure~\ref{fig:i7-8700T} left) show that
the performance per pattern improves significantly when we process
multiple patterns in one window of RF samples (which is how the experiment
was structured, since this is typical given how ATLAS systems are
usually configured). This is mostly due to the amortization of the
cost of demodulation over many patterns. The graph on the right in
Figure~\ref{fig:i7-8700T} that using 2 or 3 cores improves performance
relative to using only one core, but the improvement is far from dramatic
or linear. Using 4 or more cores actually slows the code down relative
to 2 or 3 cores. The parallelization in the CPU code is only within
FFTW and it does not appear to be particularly effective in this code,
perhaps due to the length of the FFTs.

\begin{figure}[tb]
\includegraphics[width=0.47\textwidth]{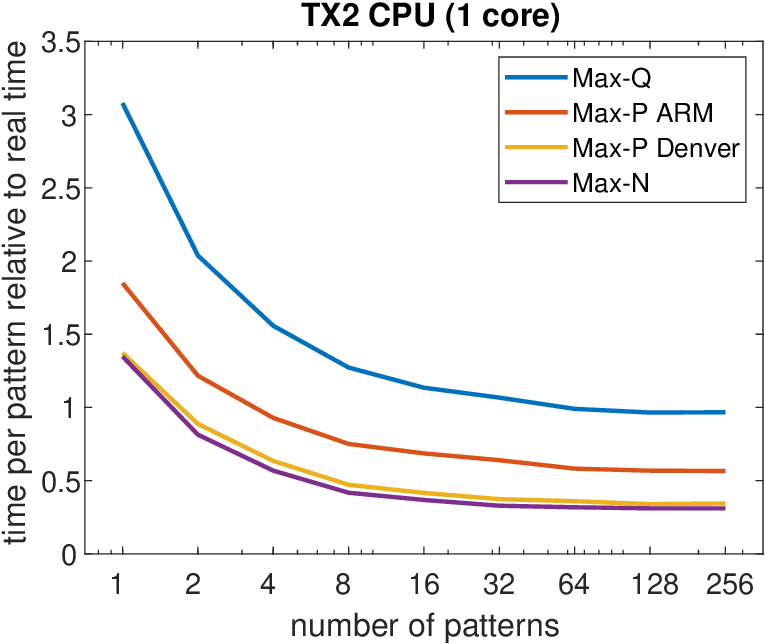}\hfill{}\includegraphics[width=0.47\textwidth]{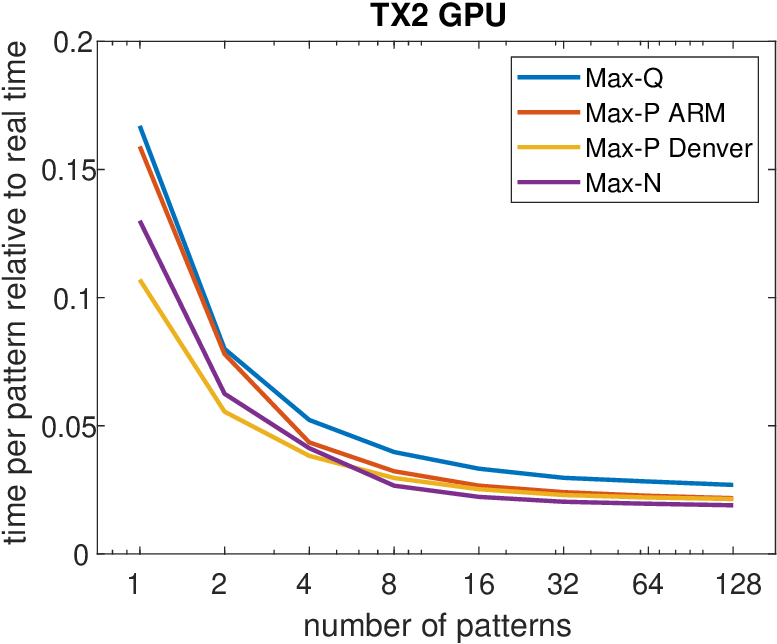}

\caption{\label{fig:searching-tx2}Searching performance on the Jetson TX2
on both the ARM cores (left) and the GPU (right) under four standard
power configurations.}
\end{figure}

Performance on the TX2 is excellent on the GPU but poor on the CPU,
as shown in Figure~\ref{fig:searching-tx2}. Our CUDA code running
on the TX is about 4.3 times faster than the single-core i7 code and
about 3 times faster than the i7 multicore runs. However, even at
the highest performance mode, the TX's CPU cores perform about 4 times
worse than the i7. 
\begin{figure}[tb]
\includegraphics[width=0.47\textwidth]{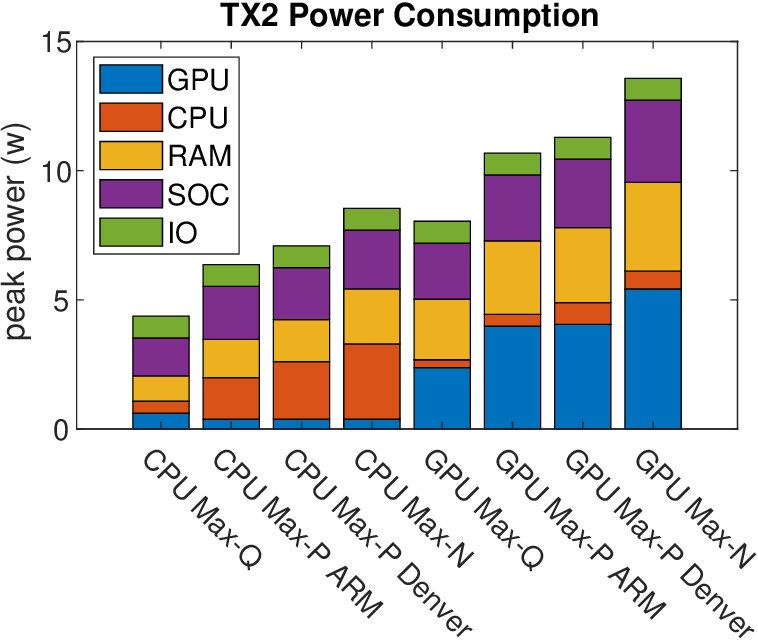}\hfill{}\includegraphics[width=0.47\textwidth]{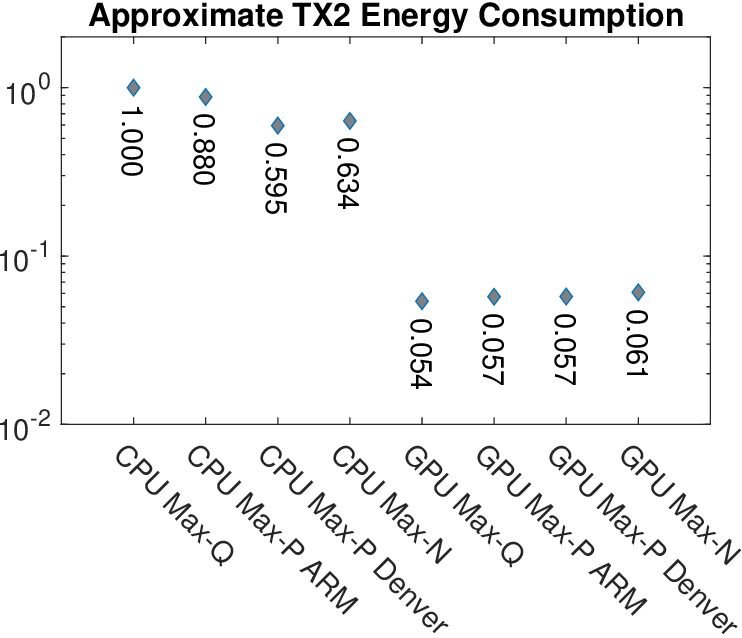}

\caption{\label{fig:tx2-power-consumption}Power consumption during searching
tasks (left), broken down by system component, and approximate total
energy consumption during searching with 16 patterns, normalized to
the largest energy expenditure (right). Both graphs show the data
for searching on either the CPU or the GPU of the Jetson TX2 and under
four standard power configurations. The rated accuracy of the power
sensors is 2\% for values above 200~mW and 15\% for smaller values.}
\end{figure}
We also measured the power consumption of the TX2 while it was running
our code. The results, shown in Figure~\ref{fig:tx2-power-consumption},
indicate that when running the GPU code, the GPU is the largest power
consumer, but the memory and other parts of the system-on-chip (most
probably the memory interface) consume a lot of power, about 50\%
of the total. The CPU and IO interfaces also consume power, but not
much. In the C-code runs, the GPU is essentially off; the CPU, memory,
and system-on-chip are the largest power consumers. The graph on the
right in Figure~\ref{fig:tx2-power-consumption} shows that the CUDA
code is about 10 times more energy efficient than the C code running
on the CPU, for the same task.

\begin{figure}[tb]
\includegraphics[width=0.47\textwidth]{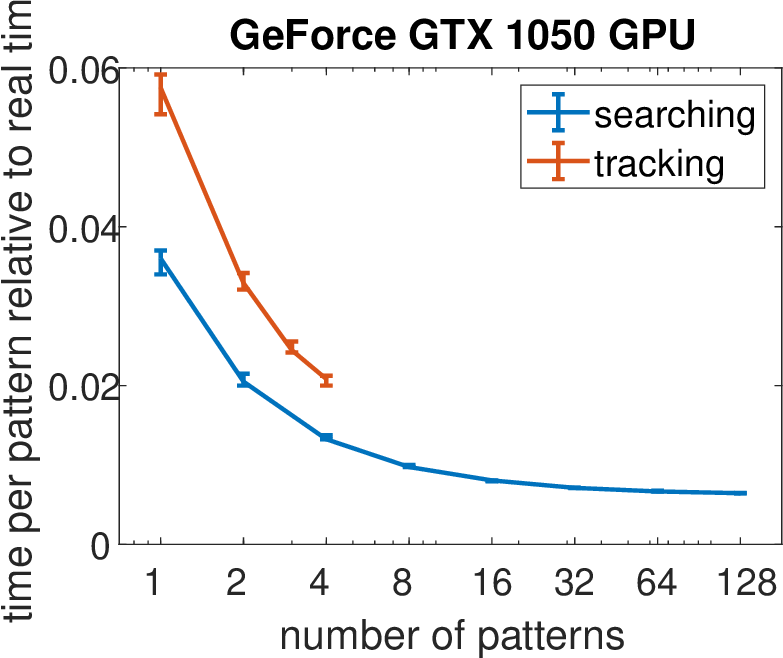}\hfill{}\includegraphics[width=0.47\textwidth]{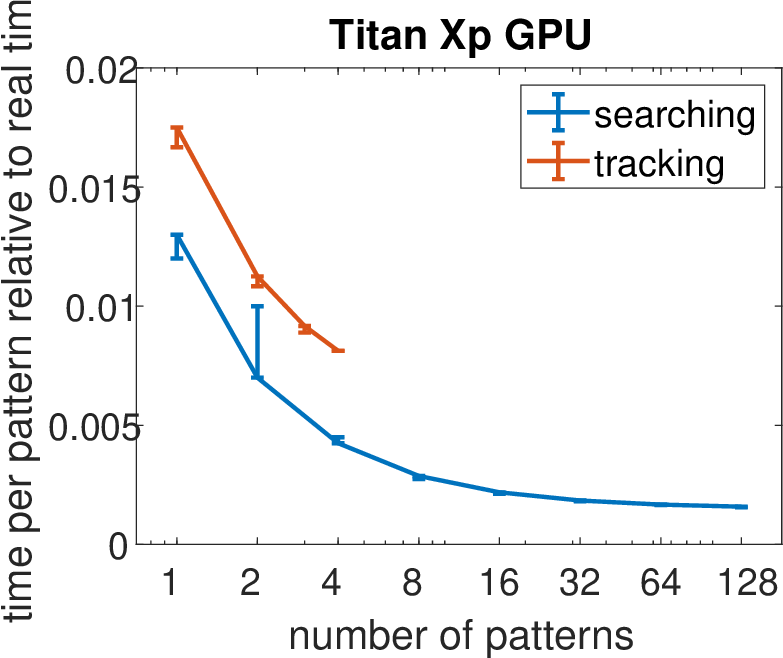}

\caption{\label{fig:gtx1050-titan-xp}The performance of two desktop GPUs,
a low-end GeForce GTX 1050 and a high-end Titan Xp, both somewhat
old. }
\end{figure}

Figure~\ref{fig:gtx1050-titan-xp} shows that our CUDA code is also
very effective on desktop and server GPUs. A low-end GPU 12.8 times
faster than a single x86\_64 desktop core that is 2 years newer. A
server GPU is 51.4 times faster than the desktop CPU.

\begin{figure}
\includegraphics[width=1\textwidth]{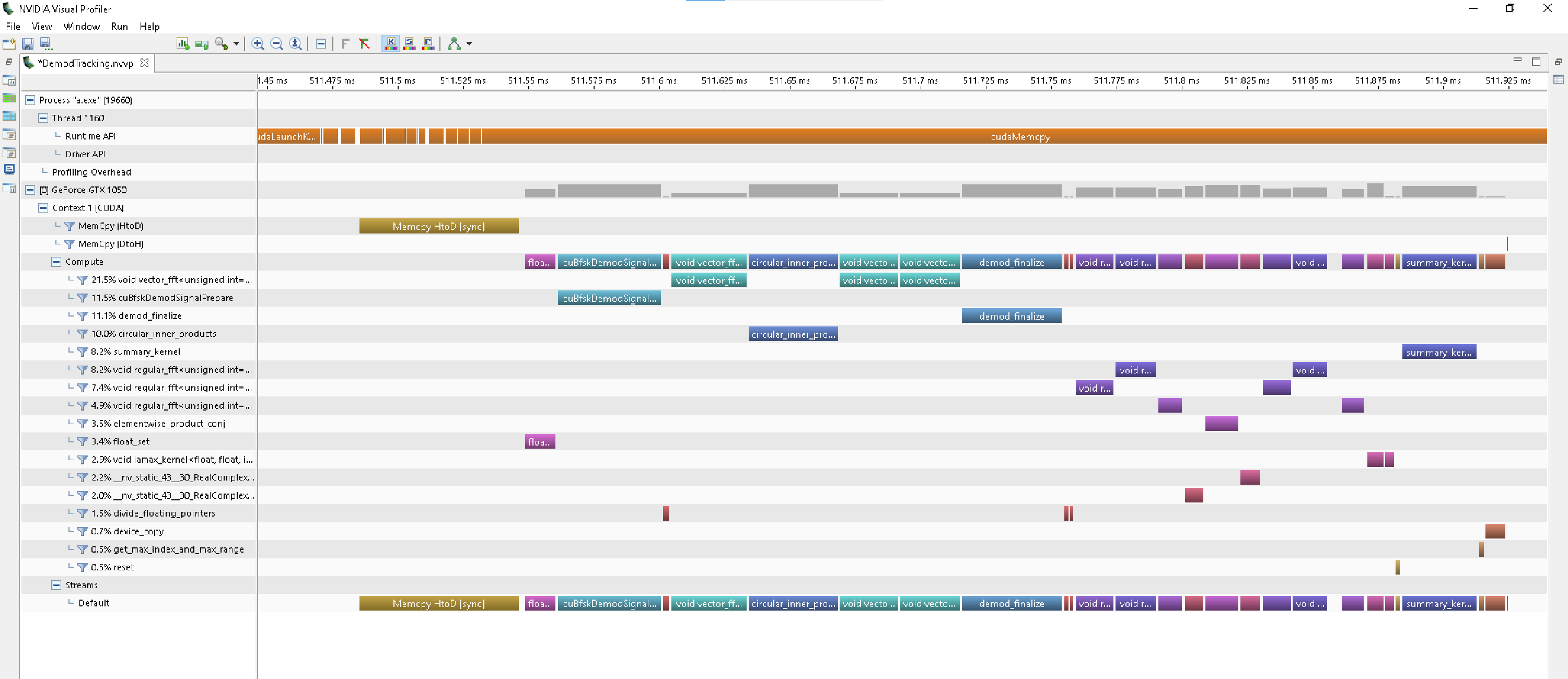}

\caption{\label{fig:CUDA-profiler}A screenshot of the CUDA profiler analyzing
a tracking task on a GTX~1050 GPU.}
\end{figure}
Figure~\ref{fig:CUDA-profiler} shows an in-depth analysis of the
performance of a tracking tag. The figure shows only the timeline
view; the profiler presents much more data not shown here. The gray
rectancles in the 6th line (denoted \emph{GeForce GTX 1050}) shows
the \emph{achieved occupancy }of each kernel.  High occupancy can
improve instruction issue efficiency as it allows to hide latency.
Rectangles that almost fill the height of the line indicate high occupancy.
This data shows that the occupancy of cuFFT is often low, around 30\%,
while kernels that we wrote, like \texttt{cuBfskDemodSignalPrepare},
achieve almost 100\% occupancy. The data also suggests that the runtime
is reasonably well balanced between kernels. That is, even if the
FFTs took no time at all, the running time of the entire computation
would not improve by more than about a factor of~2. We also see that
data transfers from and to CPU memory do not dominate the running
time. Overlapping them would improve performance but again not dramatically.
Other performance visualizations produced by the profiler and not
shown here indicate that the code exploits well the memory bandwidth
of the GPU.

\section{Deployment}

\label{sec:Deployment}Two base station (receivers) in the ATLAS system
in the Hula valley in northern Israel have recently been equipped
with computers with an NVIDIA GeForce GTX 1650 Super GPU, in order
to improve the performance of the system, especially the searching-mode
performance. Their GPUs have 1280 CUDA cores and they have 6-core
Intel i5-10500 CPUs. 

\begin{figure}
\begin{centering}
\includegraphics[width=0.75\textwidth]{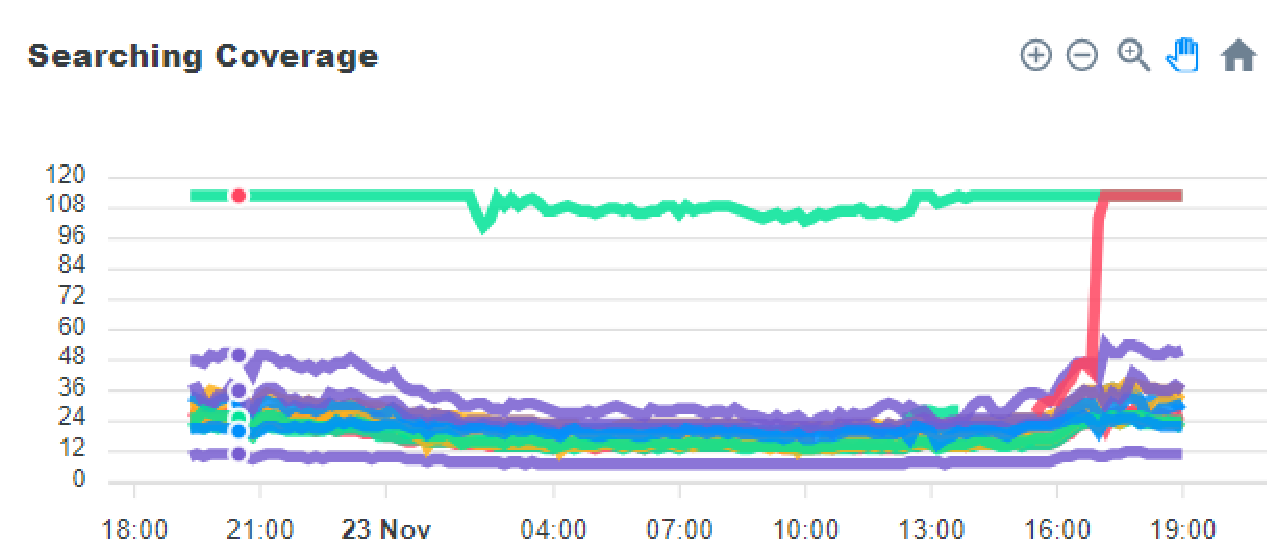}
\par\end{centering}
\caption{\label{fig:actual-searching-performance}The searching performance
of base stations in the ATLAS system in the Hula Valley during a 24-hour
period. The graph is a screenshot from a web application that is used
to monitor the system. Times are UTC and local time is 2 hours later.}

\end{figure}

Figure~\ref{fig:actual-searching-performance} shows the searching
performance of the base stations in the system over a 24-hour period
when the system was attempting to track over 100 tags. The graph shows
the fraction of RF samples processed while searching for tags. During
the night (leftmost and rightmost parts of the graph) the percentage
at base stations with GPUs is 113\%. The value is over 100\% because
of overlaps in the searching periods, which add about 13\% redundancy.
The value of 113\% implies that all the RF samples were searched for
all the tags in the searching queue. 

We see in the graph that during the day, between about 02:00 UTC and
14:00 UTC, the searching performance of all the base station dropped.
This happened because most of the tags were attached to bats, so during
the night many of them were in the tracking queue and the searching
queue was emptier; during the day, almost all the tags were in the
searching queue. At about 17:00 UTC the second GPU-equipped base station
was turned on; its performance is shown by the red curve. 

In a separate experiment, we configured the system to track 115 tags,
none of which were tracked. The searching coverage in a GPU-equipped
base station was 87\%. When we configured the base station to use
the old CPU code, performance dropped to 20\%.

\section{Related Work}

\label{sec:Related-Work}Alawieh et al.~\cite{RedFIR-FPGA-GPU-CPU}
and Hendricks et al.~\cite{RedFIR-Jetson-K1} compare the performance
of several types of compute nodes, including GPUs, CPUs, and FPGAs,
in the context of RF ToA estimation, with application to a location
estimation system called RedFIR. Their requirements are more demanding,
in the sense that RedFIR requires real-time processing of a stream
of samples, whereas we buffer samples for a few seconds and use a
priority scheduler to simplify the signal-processing code. It is well
known that real-time scheduling on GPUs is challenging~\cite{AvoidingPitfalls};
our scheduler allows ATLAS to avoid the difficulty. Also, RedFIR does
not rely on periodic transmit schedules, whereas ATLAS reduces the
computational load by tracking tags rather than just searching for
them. Finally, signal processing in RedFIR is a bit simpler than in
ATLAS because they use PSK transmitters, not FSK transmitters. Belloch
et al.~\cite{multi-gpu-expert-systems} and Kim et al.~\cite{app7111152}
present acoustic localization systems that exploit GPUs for ToA estimation.

Our use of cuFFT follows the advice of St\v{r}el\'{a}k and Filipovi\v{c}~\cite[Section~2.5]{cufft-setup-autotuning}.
More specifically, we ensure that the only prime factors of input
sizes are 2, 3, 5, and 7; we use 32-bit floating point numbers and
out-of-place transforms; we use batched transforms. Other CUDA FFT
libraries~\cite{MS-CUDA-FFT,small-FFTs-on-GPUs} appear to be no
longer maintained.

\section{Discussion and Conclusions}

\label{sec:Discussion-and-Conclusions}We have shown that by implementing
the DSP functionality of an RF time-of-arrival transmitter localization
system in CUDA, we can improve the acquisition (searching) throughput
of the system by a factor of 4 while reducing power consumption by
a factor of 5 or so relative to a baseline single-core C code, even
though the C code has been carefully optimized. Table~\ref{tab:platforms},
which summarizes the characteristics of our test platforms show that
higher-end GPUs can improve throughput dramatically higher, at the
cost of higher power consumption, and sometimes also higher cost.
The throughput of tracking modes also improves on GPU platforms.
\begin{table}[tb]

\caption{\label{tab:platforms}A comparison of GPU platforms. Column 3 shows
the number CPU and GPU cores. The 5th column shows the TDP of the
platform, either the overall power consumption or, if marked by a
$+$, of only the device itself. The cost in USD is only indicative,
and again shows either the total system cost or, when marked by a
$+$, the cost of the device. The rightmost column shows the throughput,
defined as the number of codes (tags) that can be searched for without
dropping any RF samples, assuming batches of 128 and windows of 100~ms
each; this also assumes that only 50\% of the time is devoted to searching,
the rest to tracking. The performance of the i7 processor assumes
that only one of the six cores are used.}

\centering{}%
\begin{tabular}{lcccccr}
Device & Launch & Cores & Fab & W & USD & tput\tabularnewline
\hline 
i7-8700T & Q2 2018 & $6\times\text{x86}$ & 14~nm & $35^{+}$ & 1000 & 6\tabularnewline
Jetson TX2 & Q2 2017 & $\begin{array}{c}
6\times\text{ARM}\\
+256
\end{array}$ & 16~nm & 7.5--15 & 1000 & 26\tabularnewline
GeForce GTX 1050 & Q2 2016 & $+384$ & 14~nm & $75^{+}$ & $110^{+}$ & 77\tabularnewline
Titan Xp & Q2 2017 & $+3840$ & 16~nm & $250^{+}$ & $1200^{+}$ & 315\tabularnewline
\end{tabular}
\end{table}

Our baseline code does not effectively exploit multicore CPU platforms,
even though it relies heavily on a (high-quality) parallel multicore
FFT library; this alone does not deliver good parallel speedups, perhaps
due to the modest size of the tasks. It is likely that a careful parallel
multicore implementation, perhaps in OpenMP, can improve the performance
of the C code on multicore CPUs. However, this would entail programming
that is at least as complex as our CUDA implementation, and it would
still not attain the maximum performance of the GPU code or its power-performance
ratio. 

This code is now mature and in production. The entire signal-processing
software library that ATLAS uses has been converted to CUDA (including
the PSK code). We will continue to maintain both versions and users
can switch between them easily at run time. Two base stations with
GPUs have already been deployed and they speed up the acquisition
time of the system. When all the base stations in a system have GPUs,
tracking capacity will also increase significantly. We plan to test
and deploy base station computers based on the NVIDIA Xavier AGX and/or
Xavier NX development kits (512 or 384 CUDA Volta cores, respectively,
and only up to 30W). 

\paragraph*{Acknowledgments}

Thanks to the editor and the three reviewers for comments and
suggestions that helped us improve the paper. Thanks to NVIDIA Corporation for the donation of the Jetson TX2. This
study was also supported by grants 965/15, 863/15, and 1919/19 from
the Israel Science Foundation.

\bibliographystyle{plain}
\bibliography{atlas-dsp-gpu}

\end{document}